\makeatletter
\declare@file@substitution{revtex4-1.cls}{revtex4-2.cls}
\makeatother

\documentclass[preprint2]{aastex631}

\newcommand{\tbn}{$\theta_{Bn}$ }	
\newcommand{\tbr}{$\theta_{\mathrm{BR}}$ }
\newcommand{\tbv}{$\theta_{\mathrm{BV}}$ }

\usepackage{amsmath}
\begin{document}

\title{Properties of an interplanetary shock observed at 0.07 and 0.7 Astronomical Units by Parker Solar Probe and Solar Orbiter.}

\correspondingauthor{Domenico Trotta}
\email{d.trotta@imperial.ac.uk}

\author[0000-0002-0608-8897]{Domenico Trotta}
\affiliation{The Blackett Laboratory, Department of Physics,
Imperial College London, 
London, SW7 2AZ, UK}

\author[0000-0002-7653-9147]{Andrea Larosa}
\affiliation{Department of Physics and Astronomy, 
Queen Mary University of London, 
London, E1 4NS, UK}

\author[0000-0003-3623-4928]{Georgios Nicolaou}
\affiliation{Department of Space and Climate Physics, Mullard Space Science Laboratory, University College London, Dorking, Surrey, RH5 6NT, UK}

\author[0000-0002-7572-4690]{Timothy S. Horbury}
\affiliation{The Blackett Laboratory, Department of Physics,
Imperial College London, 
London, SW7 2AZ, UK}

\author[0000-0002-6276-7771]{Lorenzo Matteini}
\affiliation{The Blackett Laboratory, Department of Physics,
Imperial College London, 
London, SW7 2AZ, UK}

\author[0000-0002-3039-1255]{Heli Hietala}
\affiliation{Department of Physics and Astronomy, 
	Queen Mary University of London, 
	London, E1 4NS, UK}

\author[0000-0001-7171-0673]{Xochitl Blanco-Cano}
\affiliation{Departamento de Ciencias Espaciales, Instituto de Geofísica, Universidad Nacional Autónoma de México, Ciudad Universitaria, Ciudad de México, Mexico}

\author[0000-0002-7419-0527]{Luca Franci}
\affiliation{The Blackett Laboratory, Department of Physics,
Imperial College London, 
London, SW7 2AZ, UK}

\author[0000-0003-4529-3620]{C. H. K Chen}
\affiliation{Department of Physics and Astronomy, 
	Queen Mary University of London, 
	London, E1 4NS, UK}

\author[0000-0002-4299-0490]{Lingling Zhao}
\affiliation{Center for Space Plasma and Aeronomic Research (CSPAR), University of Alabama in Huntsville, Huntsville, AL 35805, USA}
\affiliation{Department of Space Science, University of Alabama in Huntsville, Huntsville, AL 35899, USA}

\author[0000-0002-4642-6192]{Gary P. Zank}
\affiliation{Center for Space Plasma and Aeronomic Research (CSPAR), University of Alabama in Huntsville, Huntsville, AL 35805, USA}
\affiliation{Department of Space Science, University of Alabama in Huntsville, Huntsville, AL 35899, USA}

\author[0000-0002-0978-8127]{Christina M. S. Cohen}
\affiliation{California Institute of Technology, Pasadena, CA 91125, USA}

\author[0000-0002-1989-3596]{Stuart D. Bale}
\affil{Physics Department, University of California, Berkeley, CA 94720-7300, USA}
\affil{Space Sciences Laboratory, University of California, Berkeley, CA 94720-7450, USA}

\author[0000-0002-6577-5515]{Ronan Laker}
\affiliation{The Blackett Laboratory, Department of Physics,
Imperial College London, 
London, SW7 2AZ, UK}

\author[0000-0001-6308-1715]{Nais Fargette}
\affiliation{The Blackett Laboratory, Department of Physics,
Imperial College London, 
London, SW7 2AZ, UK}

\author[0000-0002-1296-1971]{Francesco Valentini}
\affiliation{Dipartimento di Fisica, University of Calabria, Rende 87036, Italy}

\author[0000-0001-5550-3113]{Yuri Khotyaintsev}
\affiliation{Swedish Institute of Space Physics, Uppsala, Sweden}

\author[0000-0003-0937-2655]{Rungployphan Kieokaew}
\affiliation{Institut de Recherche en Astrophysique et Planétologie, CNRS, UPS, CNES, 9 Ave. du Colonel Roche 31028 Toulouse, France}

\author[0000-0003-2409-3742]{Nour Raouafi}
\affiliation{Johns Hopkins Applied Physics Laboratory, Laurel, MD 20723, USA}

\author[0000-0001-9992-8471]{Emma Davies}
\affiliation{Austrian Space Weather Office, GeoSphere Austria, 8020 Graz, Austria}

\author[0000-0002-3298-2067]{Rami Vainio}
\affiliation{Department of Physics and Astronomy, University of Turku, FI-20014 Turku, Finland}

\author[0000-0003-3903-4649]{Nina Dresing}
\affiliation{Department of Physics and Astronomy, University of Turku, FI-20014 Turku, Finland}

\author[0000-0002-4489-8073]{Emilia Kilpua}
\affiliation{Department of Physics, University of Helsinki, Helsinki, Finland}

\author[0000-0003-1270-1616]{Tomas Karlsson}
\affiliation{KTH Royal Institute of Technology, Stockholm, Sweden}

\author[0000-0002-5982-4667]{Christopher J. Owen}
\affiliation{Department of Space and Climate Physics, Mullard Space Science Laboratory, University College London, Dorking, Surrey, RH5 6NT, UK}

\author[0000-0002-7388-173X]{Robert F. Wimmer-Schweingruber}
\affiliation{Institute of Experimental and Applied Physics, Kiel University, 24118 Kiel, Germany}




\begin{abstract}
The Parker Solar Probe (PSP) and Solar Orbiter (SolO) missions opened 
a new observational 
window in the inner heliosphere, which is finally accessible to direct measurements.
On September 05, 2022, a coronal mass ejection (CME)-driven interplanetary (IP) shock
has been observed as close as 0.07 au by PSP. 
The CME then reached SolO, which was well radially-aligned at 0.7 au, thus providing 
us with the opportunity to study the shock properties at so different 
heliocentric distances. We characterize the shock, investigate its typical 
parameters and compare its small-scale features at both locations.
Using the PSP observations, we investigate how magnetic switchbacks and ion cyclotron waves are processed upon shock crossing. We find that switchbacks preserve their 
V--B correlation while compressed upon the shock passage, and that the signature of
ion cyclotron waves disappears downstream of the shock.
By contrast, the SolO observations reveal a very structured shock transition, with
a population of shock-accelerated protons of up to about 2 MeV, 
showing irregularities in 
the shock downstream, which we correlate with solar wind structures propagating 
across the shock. At SolO, we also report the presence of
low-energy ($\sim$ 100 eV) electrons scattering due to upstream shocklets. 
This study elucidates how the local features of IP shocks and their 
environments can be very different as they propagate through the heliosphere.

\end{abstract}



\section{Introduction} 
\label{sec:intro}

Collisionless shock waves are present in a large number of astrophysical systems, and are pivotal for
efficient energy conversion and particle acceleration in our universe~\citep[e.g.,][]{Richardson2011,Bykov2019}.
Generally speaking, shocks convert directed flow energy (upstream) into heat and magnetic energy (downstream), 
and, in the collisionless case, a fraction of the available energy 
is channeled into the production of energetic particles.

Shocks in the heliosphere are unique, being directly accessible by spacecraft exploration, 
thus providing the missing
link to the remote observations of astrophysical systems~\citep{Richardson2011}. Most of our knowledge about the
\emph{in-situ} properties of collisionless shocks comes from the Earth's bow 
shock~\citep{Eastwood2015}, due to its convenient location~\citep{Dungey1979, Wilkinson2003}. Shocks besides 
the Earth's bow shock are not as well observed and understood. Interplanetary
(IP) shocks are generated as a consequence of solar activity phenomena, such as coronal mass ejections~\citep[CMEs, e.g., 
][]{Gosling1974, Webb2012, Kilpua2017} and stream interaction regions~\citep[SIR, e.g., ][]{Dessler1963, Richardson2018, PerezAlanis2023}. As elucidated in decades of
observations, IP shocks are generally weaker and show larger radii of curvature with respect to the Earth's bow
shock, thus allowing the study of collisionless shocks in highly different regimes~\citep[see][]{Kilpua2015}.

IP shocks play an important role for the
overall heliosphere energetics, due to their ability to accelerate particles to high energies and influence the
plasma environment in their surroundings \citep[see][for a review]{Reames1999}. Interestingly, the nature of such an 
interaction between IP shocks and the complex, turbulent solar wind is still largely unknown~\citep{Guo2021}. Several 
studies addressed the interaction of IP shocks with various kinds of pre-existing structures~\citep[e.g.,][]{Nakanotani2021, Zank2021, Pitna2023} as well as fully-developed turbulence~\citep{Zhao2021, Pitna2017, Zank2021}. From this point of view, important 
insights have also been provided by means of numerical modelling, looking at large-scale, MagnetoHydroDynamics (MHD) 
behaviour~\citep{Giacalone2005a, Wijsen2023} as well as the small-scale, kinetic behaviour of the shock 
system~\citep{Trotta2021, Nakanotani2022}. A complex picture is emerging, where shocks strongly influence the plasma 
environment in which they propagate, while they are strongly influenced by the self-induced and pre-existing irregularities 
they encounter~\citep{Kajdic2021, Turc2023}. Such a picture, corroborated by theoretical studies of the shock-turbulence interaction~\citep{Zank2015, Zank2021}, and has been tested in the framework of Earth's bow shock \citep{Sundberg2016,Schwartz2022}.

IP shocks have also been shown to evolve and change their behaviour at different heliocentric
distances, as elucidated by early \emph{Helios} observations in the
inner heliosphere~\citep{Lai2012}, and \emph{Ulysses} between 1
and 5 AU~\citep[e.g.,][]{Burton1992, Zhao2018}, making the study of IP shocks 
in different heliospheric environments
particularly interesting. From this point of view, the Parker Solar Probe~\citep[PSP,][]{Fox2016} 
and Solar Orbiter~\citep[SolO,][]{Muller2020} missions are opening 
a novel observational window for IP shocks at previously unexplored, small
heliocentric distances. Therefore, the early stages of shock evolution can be probed, and it is possible to
investigate how collisionless shocks influence the plasma environment close to the Sun. Examples of studies where
such new observational capabilities have been exploited are the new multi-spaceraft observations of Solar Energetic Particle events~\citep{Dresing2023}, the CMEs exhibiting ion dropouts as observed by
PSP~\citep{Giacalone2021}, and the recent crossing of a CME leg by PSP at a distance of 14 Solar
Radii~\citep{McComas2023}. A thorough investigation of the event reported in this paper, addressing the interaction between the CME and the heliospheric current sheet with a combination of remote and in-situ PSP observations is reported in ~\citet{Romeo2023}. \citet{Long2023} investigated the presence and eruption of a magnetic flux rope for this event, also combining remote-sensing and direct observations.
SolO  \emph{in-situ} measurements  were also recently used together with \emph{Wind} (at 1 au) to understand how shocks interact with and influence their environments at different
heliocentric distances~\citep[e.g.,][]{Zhao2021}. Recently, joint PSP-SolO observations elucidated the evolution of a slow plasma parcel from the solar corona to the inner heliosphere~\citep{Adhikari2022}, SolO also provided important insights into the production and dynamics of energetic
particles at IP shocks~\citep[][]{Yang2023}, exploiting the unprecedented quality of energetic particle measurements of the
Energetic Particle Detector~\citep[EPD,][]{RodriguezPacheco2020} suite.

In this work, we study a CME-driven IP shock seen by PSP at very small heliocentric distances (0.07 AU). We
characterize the shock parameters and environment, and study how the shock interacts with upstream
switchbacks (SBs), i.e., Alfv\'enic fluctuations typical of the inner heliosphere. SB observations have 
been pivotal for the PSP mission, and are emerging as a fundamental building block of the solar wind at low heliocentric 
distances~\citep[e.g.][]{Bale2019, kasper19, DudokdeWitSBs2020, nourPSPreview2023, Bale2023}, making the study of their interaction with collisionless shock waves novel and of particular interest.
The CME was later observed by SolO, at a heliocentric distance of 0.7 au. SolO was well radially-aligned with PSP, thus 
providing us with the opportunity to study the associated shock at larger heliocentric distances and therefore later evolution 
time (see Figure~\ref{fig:orbit}). The shock at SolO is characterized by a much more structured transition, and is stronger with respect to the PSP crossing.

The paper is organized as follows: in Section~\ref{sec:data} we describe the datasets used and the basic parameter estimation techniques; in Section~\ref{sec:results} we present and discuss the results, and the conclusions are reported in Section~\ref{sec:conclusions}.

\section{Data} 
\label{sec:data}

This study focuses on the analysis of \emph{in-situ} magnetic field and plasma data for PSP and SolO. At PSP, we use
magnetic field data from the FIELDS instrument~\citep{Bale2016}, while proton density, bulk
flow velocity and temperature are obtained using the Solar Probe ANalyzer-Ions~\citep[SPAN-I,][]{Livi2022}, which is part of the Solar Wind Electrons Alphas and Protons (SWEAP) investigation suite~\citep{Kasper2016}. The electron density was also estimated using the quasi-thermal
noise technique~\citep{Moncuquet2020}.

This work uses the entire \emph{in-situ} SolO instrument suite. The magnetic field is
measured by the flux-gate magnetometer~\citep[MAG,][]{Horbury2020}. Ion bulk flow speed, plasma  density and 
temperature and electron distribution 
functions are obtained from the Solar Wind Analyser suite~\citep[SWA,][]{Owen2020}. The plasma density was 
also estimated using the SolO Radio and Plasma Waves instrument~\citep[RPW,][]{Maksimovic2020}. The SolO Energetic Particle Detector~\citep[EPD,][]{RodriguezPacheco2020} is used to investigate the properties of energetic particles at SolO.  
Throughout the shock event studied here, 
we find a discrepancy between the density estimated from the moments of the distribution function 
measured by Proton and Alpha Sensor (PAS) of the SolO SWA suite and the density estimated from the 
spacecraft potential using RPW~\citep{Khotyaintsev2021} (see Figure~\ref{fig:overview}). After analysing the density computed as a moment of the electron 
distribution function~\citep[see][]{Nicolaou2021}, we decided to use the density estimated from RPW for our analysis.

\section{Results} \label{sec:results}
\begin{figure}
	\includegraphics[width=.49\textwidth]{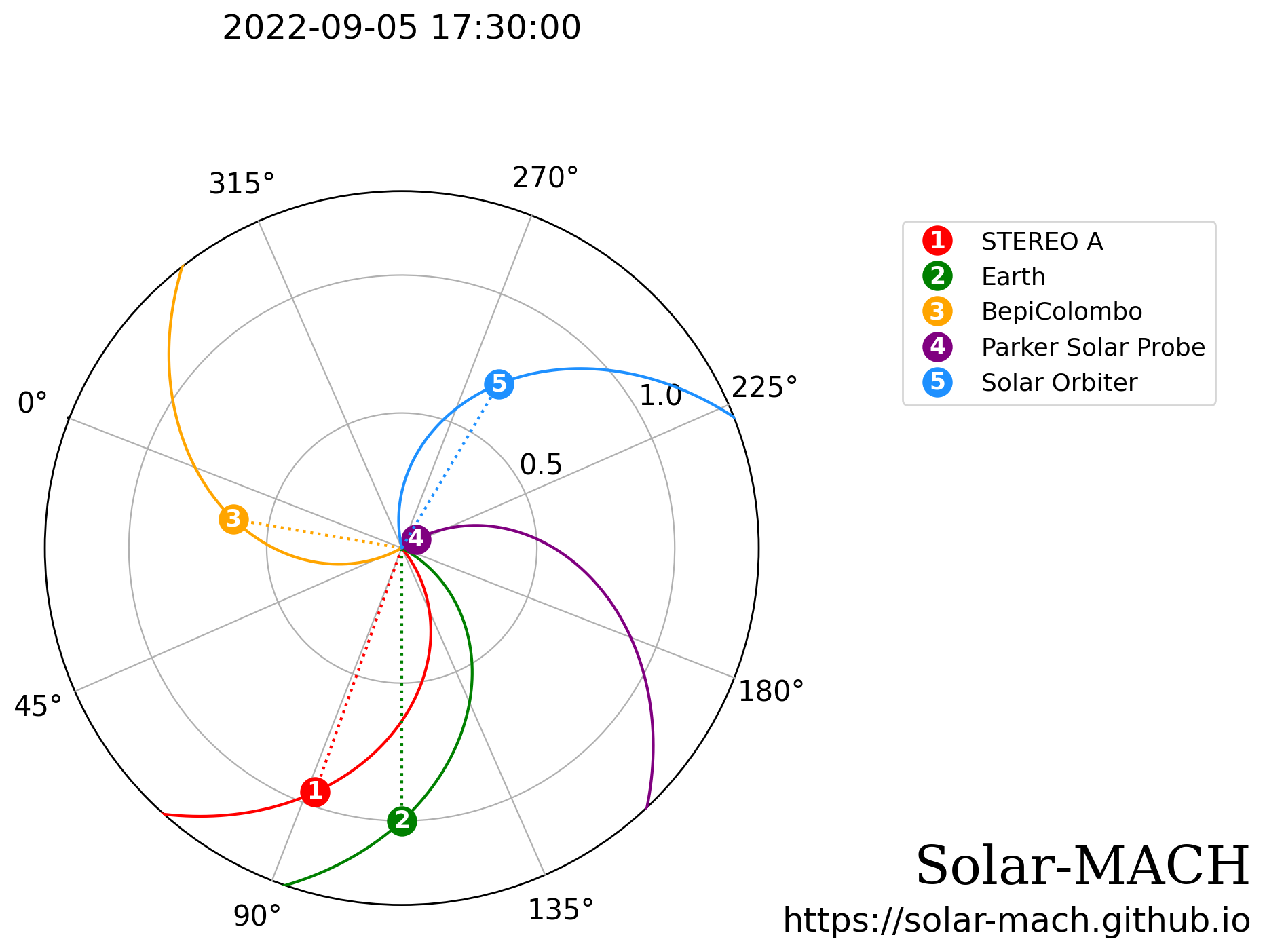}
    \caption{Spacecraft configuration at 17:30 UT of Sep. 05, 2022 (immediately after the IP shock crossed PSP). In this plot (generated using the Solar MAgnetic Connection HAUS tool ~\citep[Solar-MACH,][]{Gieseler2023MACH}), the Sun is at the center, the Earth is represented by the green circle, and the dashed (solid) lines indicate the spacecraft radial (along a Parker Spiral with a nominal speed of 400 km/s) connection.}
    \label{fig:orbit}
\end{figure}

\subsection{Event Overview} \label{subsec:event_overview}
On September 5, 2022, a CME erupted from the Sun and into 
interplanetary space. The CME was not Earth directed as the eruption happened on the far side of the Sun. An overview of the orbital configuration is shown in Figure~\ref{fig:orbit}.

\begin{figure*}
	\includegraphics[width=\textwidth]{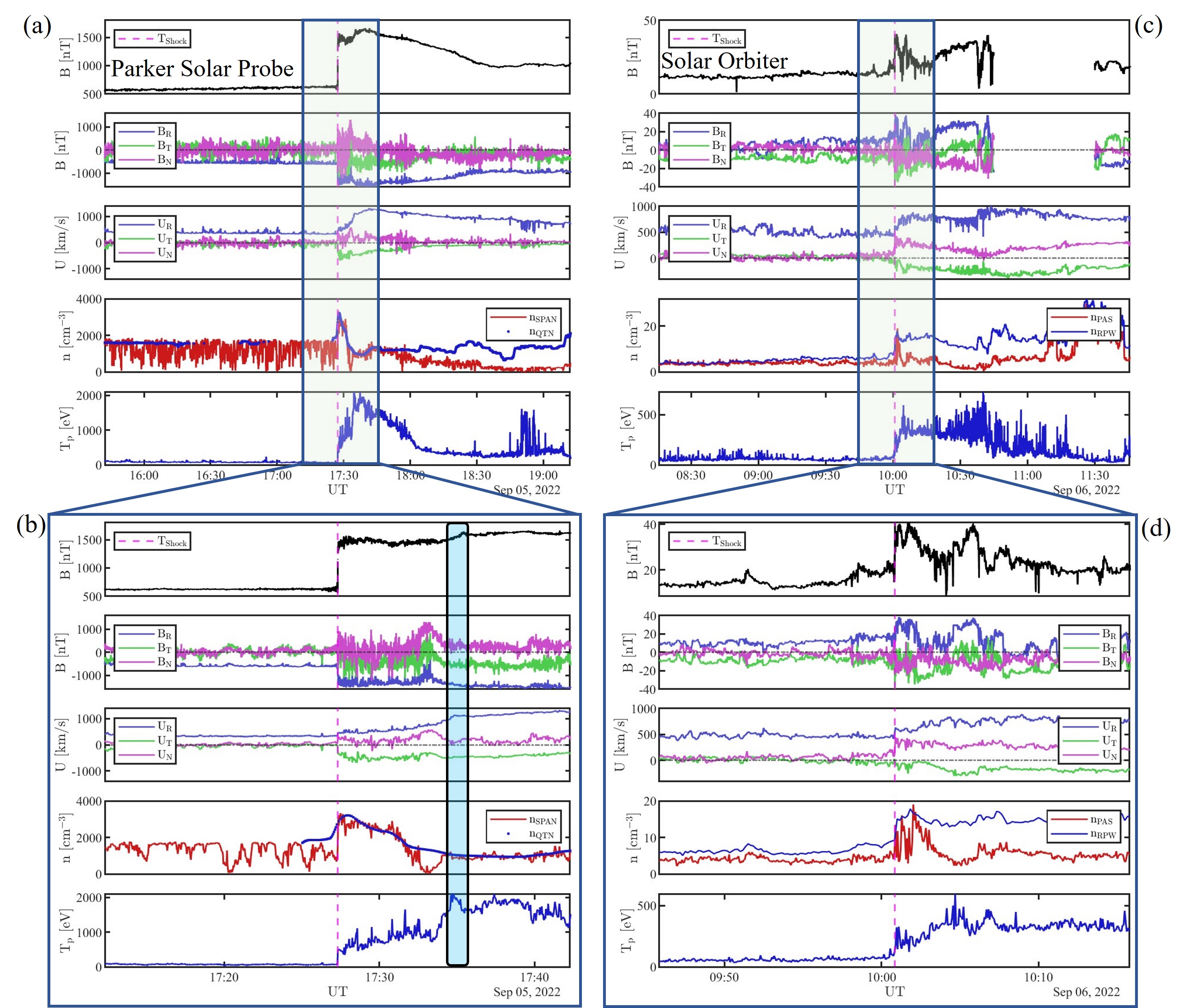}
    \caption{Magnetic field magnitude, its components in the RTN system, ion bulk flow velocity, density and temperature as measured by PSP (left) and later on by SolO (right). The plots show an overview of the event (a,c) with a zoom around the shock transition at both spacecraft (b,d). The dashed magenta line marks the shock transition, and the blue shaded area in panel (b) indicates the approximate end of the CME sheath as observed by PSP.}
    \label{fig:overview}
\end{figure*}

A fast-forward shock, driven by the CME, was detected by PSP as close as 0.07 AU ($\sim$ 15 $\mathrm R_\sun$) at 17:27:19 
UT, making this observation the closest to the Sun to date. An overview of the \emph{in-situ} magnetic field and plasma 
quantities is shown on the left-hand side of Figure~\ref{fig:overview}. A sharp transition in magnetic field, with a jump 
from about 600 to about 1600 nT occurs at the shock. The velocity profile across the shock transition shows two interesting features, 
namely a large flow speed deflection in the transverse direction $\mathrm{V_T}$, and a steady rise in $\mathrm{V_R}$ 
further downstream, up to about 17:35. We have marked this time as 
the end of the downstream (sheath) region of the event, i.e., the point where the leading edge of the CME flux rope likely starts~\citep[see also][]{Romeo2023}. However, we note that the boundary between the sheath and CME flux rope is not clear for this event. 
The properties of the CME sheath
at such low heliocentric distances (and therefore early evolutionary stages) are very interesting and will be object of future work, comparing them with substructures and properties of CMEs at larger heliocentric distances~\citep{Kilpua2017}.

\begin{table*}
	\centering
	\caption{In-situ shock parameters at PSP (left) and SolO (right). In the table are displayed,  from top to bottom, the shock arrival time, Carrington longitude and latitude at the time of shock arrival, spacecraft heliocentric distance at the time of shock arrival, shock normal vector in $\mathrm{RTN}$ coordinates, shock normal angle $\langle \theta_{Bn} \rangle$, magnetic compression ratio $\langle \rm{r}_B \rangle$, gas compression ratio $\langle \rm{r} \rangle$, shock speed $v_{\mathrm{sh}}$, upstream plasma beta $\beta_{\mathrm{up}}$, fast magnetosonic and Alfv\'enic Mach numbers ($\rm{M_{\rm{fms}}}$ and $\rm{M_{\rm{A}}}$, respectively). The shock speed $v_{\mathrm{sh}}$ is along the shock normal and computed in the spacecraft rest frame using the mass flux conservation law.}
	\label{tab:tab_event}
	\begin{tabular}{|l|c|r|} 
		\hline
        {}& Parker Solar Probe & Solar Orbiter \\
        \hline
        Shock Time [UT] & 05 Sep 2022, 17:27:19 & 06 Sep 2022, 10:00:51  \\        
        Carrington longitude [$^\circ$] & 232 &  251.9 \\
        Carrington latitude [$^\circ$] & -1.8 & -3.6  \\
        Heliocentric dist. [AU] & 0.07 & 0.7 \\
        $\langle \hat{\mathrm{n}}_{\mathrm{RTN}} \rangle$ & [0.5 -0.8 0.2] & [0.6 -0.2 0.7] \\
        $\langle \theta_{Bn}\rangle$ [$^\circ$] & 53 & 51 \\
        $\langle \rm{r}_B \rangle$ &  2.3 &  1.9 \\
        $\langle \rm{r} \rangle$  &  1.6 &  2 \\
        $\langle v_{\rm{sh}} \rangle$ [km/s] &  1520 &  942 \\
        $\beta_{\mathrm{up}}$ & 0.1 & 0.6 \\
        $\rm{M_{\rm{fms}}}$ & 3.8 & 3.2 \\
        $\rm{M_{\rm{A}}}$ & 3.9 & 3.8 \\
		\hline
		\hline
	\end{tabular}
\end{table*}
We carried out a comprehensive characterization of the shock parameters locally observed by PSP. Given the density drops 
visible in the upstream region in Figure~\ref{fig:overview}(b), due to the core of the ion distribution function not being well-inside 
the SPAN-i field of view~\citep{Livi2022}, we estimated the density using the spacecraft quasi-thermal noise (QTN) for our 
shock analysis. Using a systematic collection of averaging windows spanning a few seconds to three minutes (with the 
technique described  in~\citet{Trotta2022b}) upstream/downstream of the shock, we used the 
Mixed Mode 3 method~\citep[MX3][]{Paschmann2000} to estimate the average shock normal $\langle n \rangle = [0.5, -0.8, 
0.2]$, though similar values are obtained with the other 
MX  methods as well as magnetic coplanarity. Propagation direction deviating clearly from the radial direction is comptatible with the picture in which PSP is crossing a flank of the CME event. Our estimation of \tbn, i.e., the angle between the shock normal and the upstream magnetic field, a crucial parameter influencing shock behaviour, reveals that we are in the presence 
of an oblique (\tbn $\sim 53^\circ$), supercritical shock with moderate Alfv\'enic and fast magnetosonic Mach numbers ($\mathrm{M_A \simeq 3.9\,,\, M_{fms} \simeq 3.8}$, respectively) with 
respect to other IP shocks observed at or near 1 au~\citep{Kilpua2015}. 
Furthermore, we find that the shock propagates at the very high speed of about 1500 km/s in the 
spacecraft frame and along the estimated shock normal. At PSP, upstream of the shock the magnetic field is mostly radial and interspersed with one-sided Alfv\'enic fluctuations \citep{gosling2009ApJ...695L.213G} of the magnetic field and ion bulk velocity, known as magnetic 
switchbacks, with angular deflections in the magnetic field of moderate amplitude, as expected for very low 
heliocentric distances~\citep[e.g.][]{Jagarlamudi_2023}. SBs are a crucial feature of the solar wind in the inner 
heliosphere, as extensively shown in previous literature~\citep[]{DudokdeWitSBs2020, volodiaSBs2020ApJ...893...93K, larosa2021A&A, Liang2021, pecoraSBs2022ApJ, LiuSBs2023ApJ,  Jagarlamudi_2023, nourPSPreview2023}. Here, we have the opportunity to study how such structures are processed by shock waves, a crucial point of this work summarized in Section~\ref{subsec:shockback}.

\begin{figure*}
	\includegraphics[width=\textwidth]{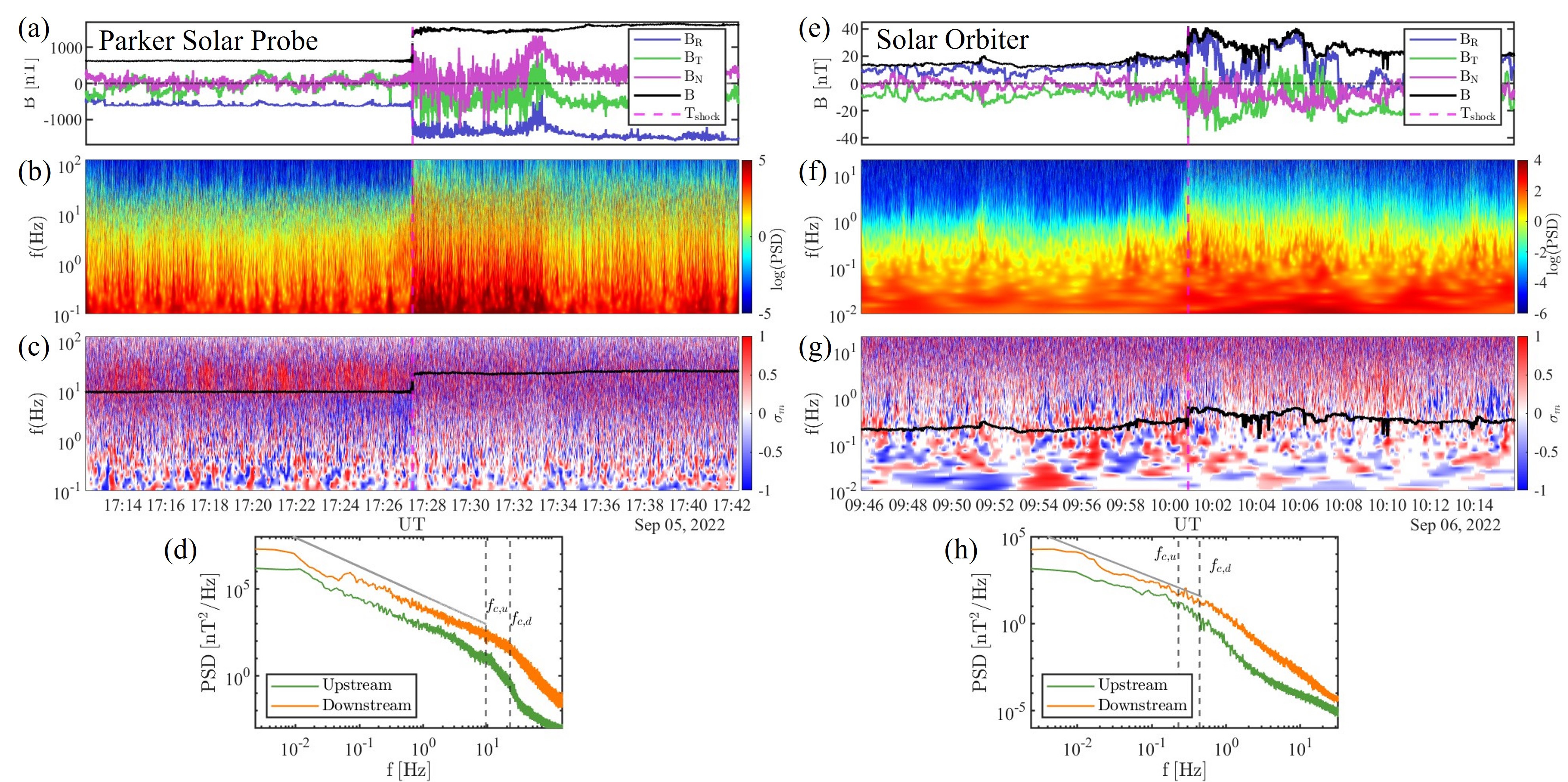}
    \caption{Magnetic fluctuations in the shock environment at PSP (left) and SolO (right). The Figure shows, from top to bottom: magnetic field magnitude and its components, trace wavelet spectrogram of the magnetic field (colormap), the normalised magnetic helicity $\sigma_m$ spectrogram, with the proton cyclotron frequency in the interval (black line), magnetic field power spectral densities (PSDs) computed in the 10 minutes upstream (green) and downstream (orange) of the shock.}
    \label{fig:wav_turb}
\end{figure*}

At SolO (Fig.~\ref{fig:overview} right), the shock crossing appears much more structured compared 
to the PSP observation. Strong 
transverse flow deflections are still present, with a strong increase in 
$\mathrm{V_N}$ at the shock transition. The plasma density, as measured 
from the SWA-PAS and RPW instruments, yields two different values. For the shock 
parameter estimation, we used the density as measured by the RPW, since it is in agreement with
the value obtained from EAS electron moments (Figure~\ref{fig:eas_shocklets}). However, locally
computed shock parameters do not change dramatically when using the density
derived from PAS.
Steep upstream enhancements of magnetic 
field magnitude are 
found ahead of the shock (09:52, 09:56 UT), compatible with very rarely observed 
shocklets at strong IP 
shocks~\citep{Wilson2009, Trotta2023a}.
A high degree of magnetic field structuring is also found 
downstream of the shock, indicating a high level of complexity 
for this shock 
crossing. At SolO, the shock parameters estimation shows values very close 
to the ones observed at PSP (see Table~\ref{tab:tab_event}). However, such values 
are obtained using very local averaging windows for the
upstream/downstream quantities ($\sim$ 10 seconds to 1 minute). The overall 
behaviour of the shock at SolO indicates a high level of variability. 
Indeed, using larger averaging 
windows (about 6 to 10 minutes up/downstream), we find parameters 
compatible with a quasi-parallel 
shock transition (\tbn $\sim 30^\circ$) with high
Mach numbers ($\mathrm{M_A \sim M_{fms} \sim 7}$). It 
is worth mentioning that, in addition to the local variability examined here,
a different behaviour of 
the two shocks at PSP and SolO is expected, due to the crossing of the
event at two different points in 
space and time. The longitudinal separation of PSP and SolO is of about 20$^\circ$, and depending on the CME width, the shock will be crossed in different locations. However, such a comparison of shock parameters, summarized in 
Table~\ref{tab:tab_event}, is useful when addressing the evolution of the
whole event, and it may be 
important to support remote-sensing observations of the event as well as modelling of the CME evolution.

It is interesting, at this point, to characterize the environment in which the shock at PSP and SolO 
are propagating and how it is influenced by the shock passage. To this end, we studied the magnetic 
fluctuations in the $\sim$ 15 minutes before/after the shock arrival, with the relevant analyses shown 
in Figure~\ref{fig:wav_turb}.

\begin{figure*}
	\includegraphics[width=\textwidth]{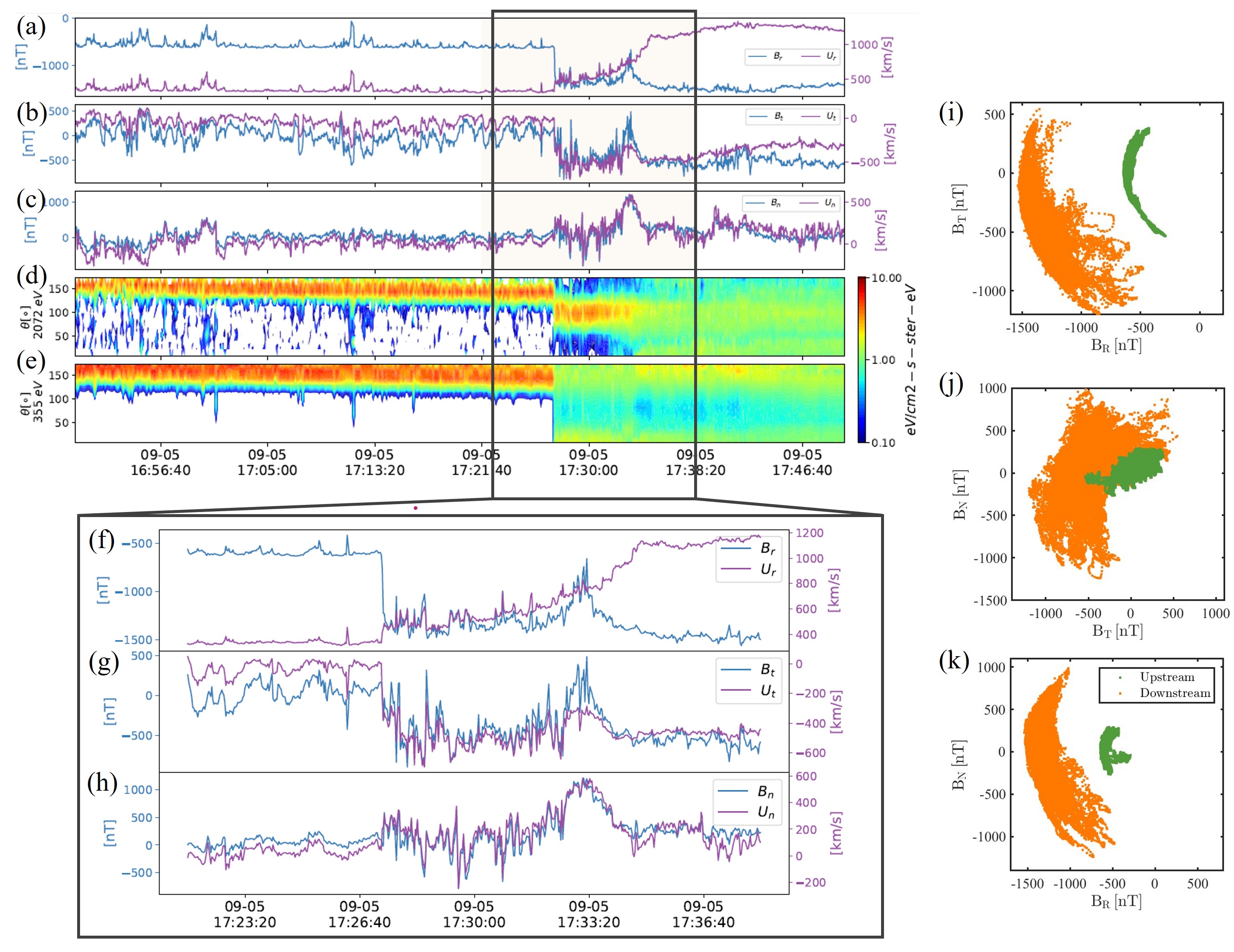}
    \caption{PSP observations of magnetic switchbacks transmission across the shock. (a) to (c): magnetic field (blue) and ion bulk flow velocity (plum) components. (d) and (e): pitch angle distributions for electrons with 2072 and 355 eV energy, respectively. (f) to (h): zoom around the shock crossing (shaded area in top panels). Scatterplots of radial and tangential (i), tangential and normal (j) and radial to normal (k) components of the magnetic field for the 4 minutes upstream of the shock crossing (green) and 4 minutes downstream of it (orange).}
    \label{fig:shockback1}
\end{figure*}

The shock crossing at PSP is characterized by a sharp transition, with no signs of upstream wave foreshock,
as elucidated by the trace wavelet spectrogram in Figure~\ref{fig:wav_turb}(b). This is compatible with other observations 
of IP shocks, particularly at such oblique geometries~\citep[e.g.][]{BlancoCano2016}. It is worth noting that micro-instabilities of the shock front happening at ion kinetic scales, such as rippling~\citep{Trotta2023b} may still be present, albeit not well resolved by PSP. The level of magnetic fluctuations 
is enhanced in the downstream sheath region, lasting about 6 minutes. Such an enhancement in magnetic fluctuations is further shown
in Figure~\ref{fig:wav_turb}(d), where the power spectral density in the 5 minutes upstream/downstream of the shock has been
computed (green and orange lines, respectively). Here, it can be seen that the level of fluctuations increases by a factor of 4, and 
the downstream spectra show a flattening around the ion cyclotron frequency, a behaviour compatible with the spectral behaviour of 
IP shocks~\citep{Pitna2021}, with some observations of turbulence in the Earth's bow shock 
environment~\citep[e.g.][]{Sahraoui2020} and also compatible with modelling of turbulence transmission across shocks~\citep[see Fig. 18 in][]{Zank2021}. However, because the downstream flow speed is much larger than the upstream, and the shock changes the plasma parameters abruptly, the level
	of fluctuations may be overestimated. While the increase in the level of the frequency spectrum has been documented in previous literature~\citep{Zhao2021, Park2023}, it remains to be understood if IP shocks inject new turbulence or they simply modify the background plasma properties. However, preliminary analysis shows that the spectral break happens around the ion skin depth $d_i$, compatible with statistical work carried out by \citet{Park2023}. Further studies on the matter, relevant also for understanding the behaviour of turbulence at very low heliocentric distances~\citep{Zank2020slab} are further complicated by the geometrical constraints imposed by the single-spacecraft nature of the observations. These will be the objects of future work. 

\begin{figure*}
	\includegraphics[width=\textwidth]{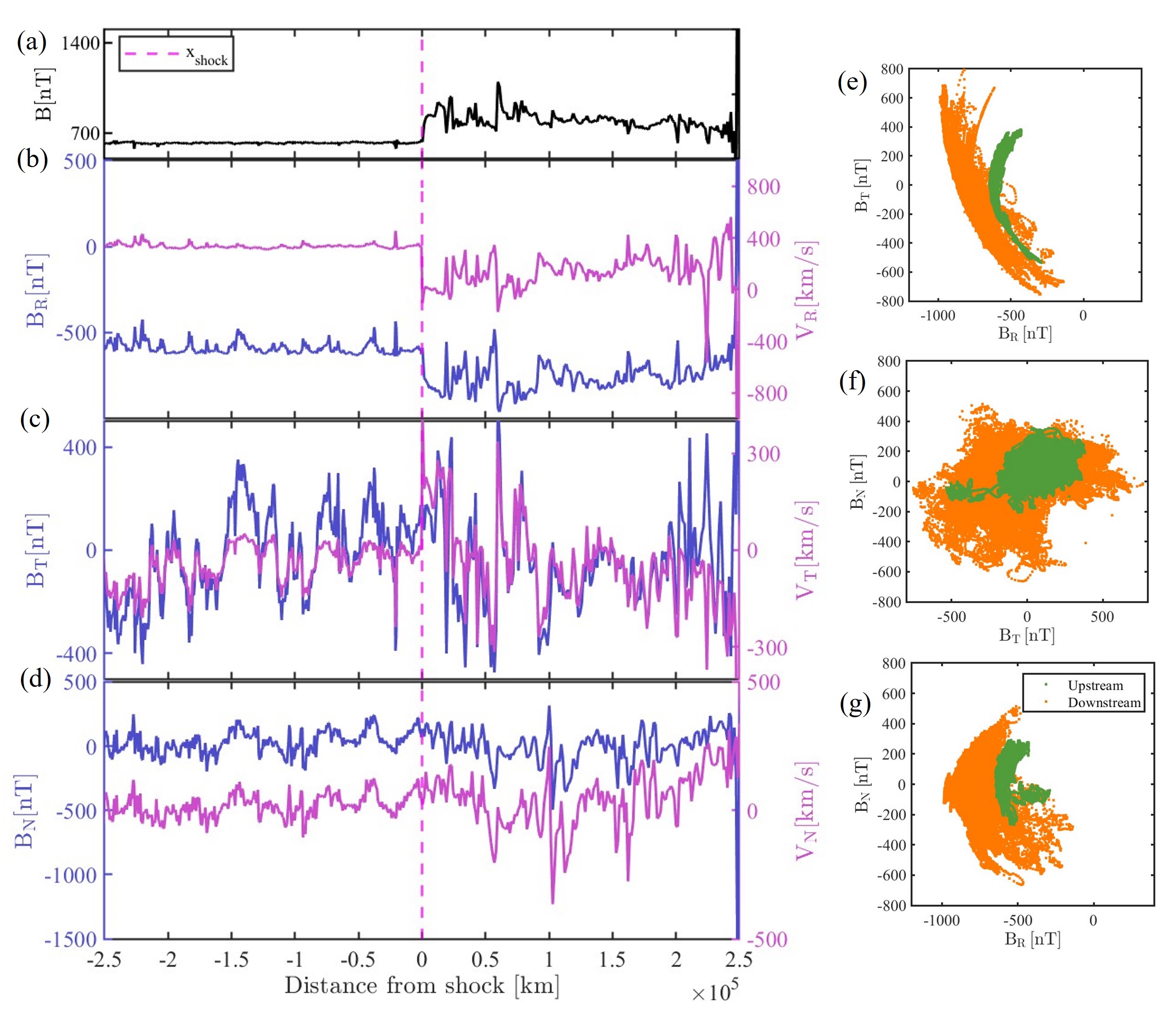}
    \caption{PSP observations of: (a)-(d): Magnetic field magnitude (black), magnetic field components (blue) and ion bulk flow speed components (plum) as a function of distance from the shock, where the downstream quantities have been decompressed using the Rankine-Hugoniot relations with the average shock parameters reported in Table~\ref{tab:tab_event}. (e)-(g): Magnetic field excursion scatter plots as in Figure~\ref{fig:shockback1}, performed on the decompressed quantities. Note that the scales in panels (e)-(g) are different from the ones used in Figure~\ref{fig:shockback1}.}
    \label{fig:decompression}
\end{figure*}

To further investigate the wave properties across the shock, we computed the normalised magnetic helicity, 
defined as 
\begin{equation}
\sigma_m = \frac{2 \Im \left(\tilde{B_T^{\star}}\tilde{B_N}\right)}{|\tilde{B_R}|^2 + |\tilde{B_T}|^2 + 
|\tilde{B_N}|^2},
\end{equation}
 where $\mathrm B$ indicates the magnetic field components, the $\tilde{}$ represents the wavelet-
transformed quantities and $\star$ represents the complex conjugation operation~\citep{Matthaeus1982}. 
Upstream of the shock at PSP, we observe a clear signature of consistently high $\sigma_m$ at ion scales,
compatible with ion cyclotron wave activity (see the red patches by the ion plasma frequency line in 
Figure~\ref{fig:wav_turb}(c)). Ion cyclotron waves, crucial components of the solar wind as elucidated by PSP 
observations~\citep{Telloni2019, Verniero2020ApJS}, have been shown to be very important for energy dissipation and solar wind 
heating~\citep[e.g.][]{Woodham2018,Bowen2022}. In our observations, we note that the magnetic helicity signature of ion cyclotron waves 
found upstream of the shock is lost in the shock downstream, as discussed in Section~\ref{subsec:shockback}, together 
with an explanation of why such a behaviour is observed.

The magnetic fluctuations environment at SolO is different than that found at PSP (right-hand side of 
Figure~\ref{fig:wav_turb}). Here, the shock transition appears much more complex, especially due to the presence of larger scale structures at the SolO shock with respect to the PSP one. In the wavelet spectrum of 
the magnetic field, enhanced power extending to small ($\sim$ 1 Hz) scales is found corresponding to the upstream 
shocklet activity. The downstream appears populated with strong compressive and non compressive magnetic field 
fluctuations, indicative of the fact that the shock propagated through a very structured portion of the solar 
wind, as it can be noted from the magnetic field behaviour in the 10 minutes downstream of the shock~\citep{BlancoCano2019,Kropotina2021}. The spectral behaviour of turbulence in being transmitted from the shock 
upstream to downstream is similar to the one observed for the shock at PSP, and compatible with previous multi-
spacecraft studies of turbulence processed by IP shocks~\citep{Zhao2021}. 
It is interesting to note that, in the IP 
case, an inertial range is always recovered downstream of IP shocks, 
in contrast with some observations of Earth's magnetosheath where an inertial range is not observed downstream of 
Earth's bow shock~\citep{Sahraoui2020}, often interpreted as the Earth's bow shock ``resetting'' the turbulent cascade. Finally, we note that some features of ion-cyclotron wave activity are seen 
upstream of the SolO event, though they are much less clear than the PSP observation.

\subsection{The shock at Parker Solar Probe} \label{subsec:shockback}

\begin{figure}
	\includegraphics[width=.5\textwidth]{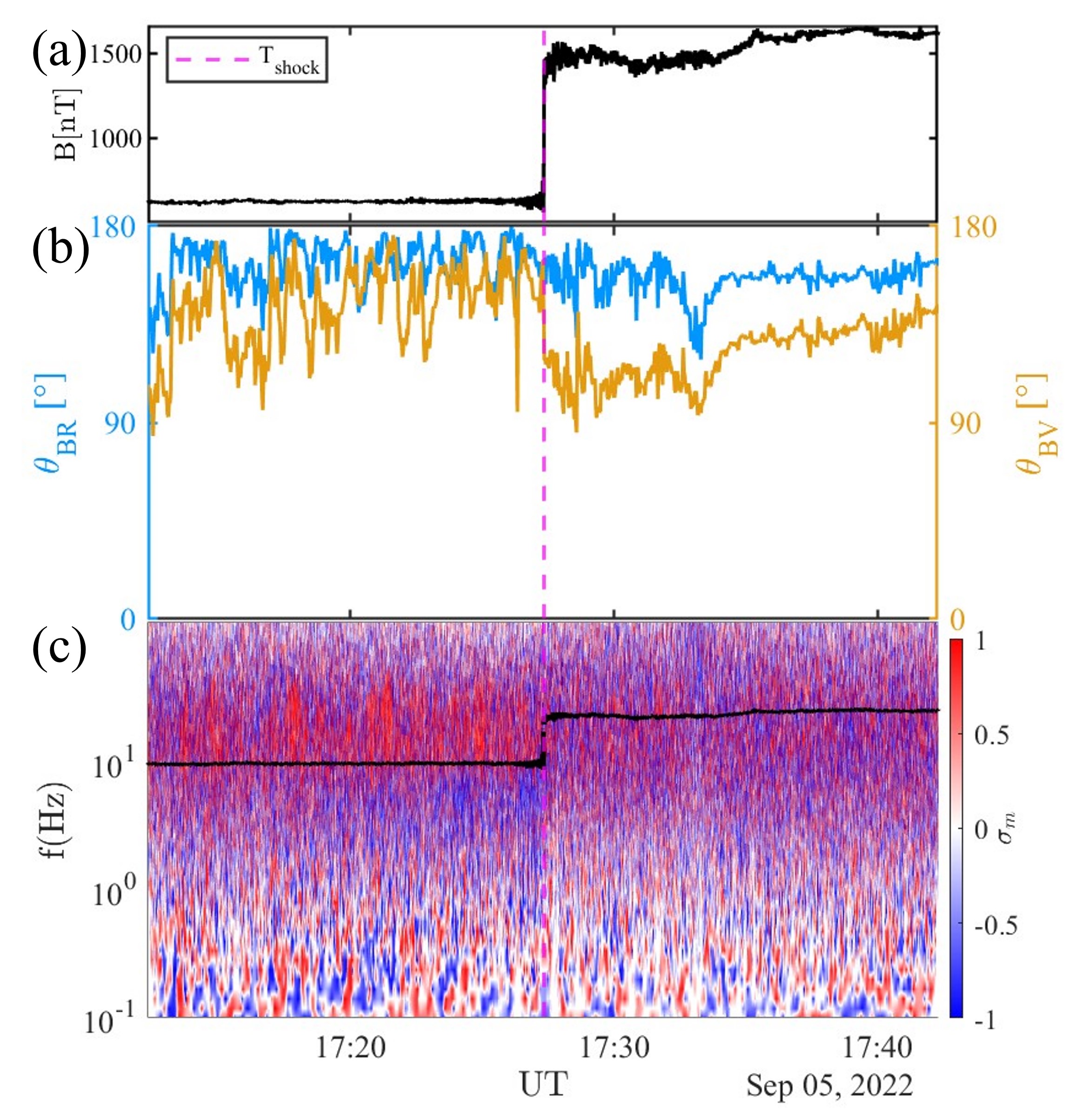}
    \caption{The IP shock interaction with ion cyclotron waves. (a): magnetic field magnitude around the shock crossing by PSP. (b) Angle between local magnetic field and radial direction (\tbr, blue) and ion velocity (\tbv, yellow). (c): Reduced magnetic helicity for the interval (as in Figure~\ref{fig:wav_turb}).}
    \label{fig:ion_cyclotron}
\end{figure}

In this section, we focus on the PSP observations of the shock interacting with 
pre-existing fluctuations in the inner-heliospheric solar wind and its features. 
Particular interest is focused on how switchbacks (SBs) are processed by the IP shock wave. 

In Figure~\ref{fig:shockback1} we observe the high degree of correlation between 
the magnetic and velocity fields. The presence of moderate amplitude SBs and their 
one-sided nature~\citep{gosling2009ApJ...695L.213G}, is evident especially in the 
$\mathrm{B_R}$ and $\mathrm{V_R}$ components
(Figure~\ref{fig:shockback1}(a)) upstream of the shock. 
In the downstream (sheath) region, the high degree of correlation is preserved, as 
elucidated by the Figure zoom (panels f, g, h). These observations 
reveal that the Alfv\'enic nature of the SBs 
is preserved downstream of the shock. However, the downstream SBs present larger, 
shock-driven variations in the field magnitude. Small deviations from a 
constant B state have been already observed at SBs boundaries 
\citep{volodiaSBs2020ApJ...893...93K,Farrell2020ApJS, larosa2021A&A}. Here we show a different compressive effect, resulting from the propagation across the IP shock. As discussed in 
Section~\ref{subsec:event_overview}, the downstream region is characterized 
by stronger fluctuations driven by the shock, including small-scale, 
compressive fluctuations that are not Alfv\'enic.


From this point of view, it is interesting to note 
that the SBs signature is not completely lost upon shock passage, but rather the SBs 
are processed by the shock. Considering the Rankine-Hugoniot relations describing 
the change in plasma parameters across MHD shocks~\citep[e.g.][]{Krehl2015}, we 
expect, for the SBs, a compression along the shock 
normal direction, and a stretch along the two shock transverse directions. The extremely short duration of the sheath poses a strong limitation on a more quantitative study of how SB are affected by the shock. For instance, the Z parameter would be helpful in addressing such an issue~\citep{DudokdeWitSBs2020}. Such an investigation will be carried out in future works focusing on several shocks at low heliocentric distances.

To further investigate how the shock affects the upstream SBs, we study the 
magnetic field excursion in the regions upstream and downstream of the shock 
(Figure~\ref{fig:shockback1}(i-m)), as done in other studies elucidating the nature 
of fluctuations in the solar wind~\citep[e.g.,][]{Matteini2015}.
Two effects arise in the magnetic field fluctuations after the shock crossing. First of all, due the change in background field magnitude, the arc 
of polarization upon which the magnetic field fluctuates become wider, as 
particularly evident in the $\mathrm{B_R}$-$\mathrm{B_T}$ scatter plot 
(Figure~\ref{fig:shockback1}(i)). It is also possible to observe that
downstream magnetic field fluctuations cover a larger area in the magnetic field 
excursion space, due to the presence of compressive fluctuations
in the shock downstream. From Figure~\ref{fig:shockback1}(i-j-k), it is possible to note that the arc of polarization is displaced from the R-T plane into both the R-T and R-N 
plane when comparing the shock upstream to the downstream, consistent with compression along the normal and the stretching along the perpendicular directions. The shock-SBs interaction may have important consequences for the SBs' ability
to propagate without dispersive effects, and therefore it may affect the SBs 
lifetime, since the constant B magnitude is a necessary 
condition for the unperturbed propagation of large amplitude Alfv\'en waves \citep{barnesHollweg1974}.

\begin{figure}
	\includegraphics[width=.5\textwidth]{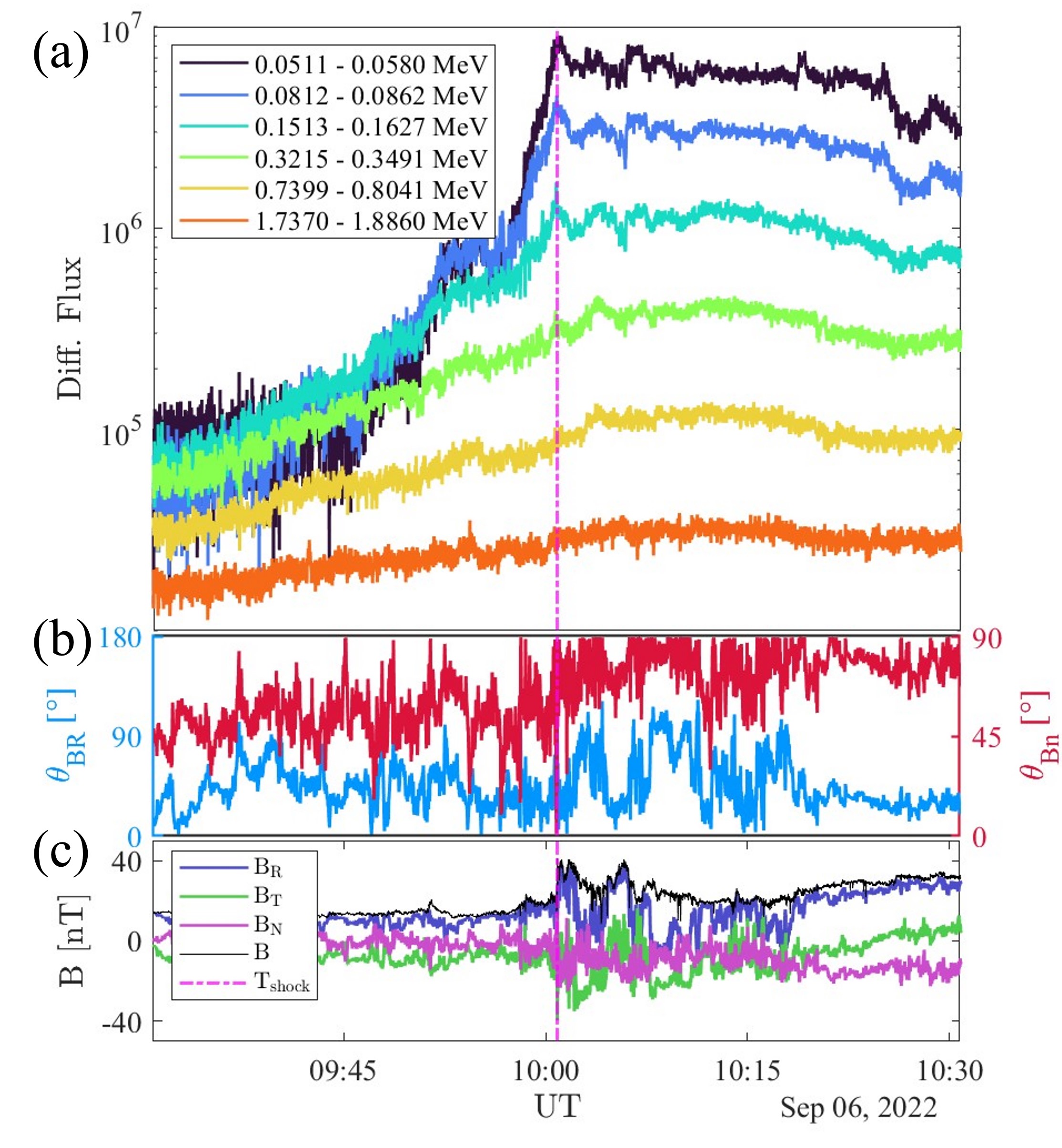}
    \caption{(a): Energetic ion fluxes in the $\sim$0.05 - 2 MeV energy range as observed by the Sun telescope of EPD-EPT instrument of Solar Orbiter. (b): Local \tbr (blue) and \tbn (red) angles. (c) Magnetic field magnitude (black) and its components. The dashed magenta line indicates the shock crossing time.}
    \label{fig:solo_tbn}
\end{figure}

The above assessment of the SBs features upstream/downstream of the shock suffers from limitations related to the single-spacecraft nature of the observations. First, it is known that the features of SBs (deflections, duration) tend to have broad distributions~\citep[e.g.,][]{larosa2021A&A}, making it difficult to estimate the extent to which the processed, downstream SBs were comparable to their upstream, unshocked counterparts before they crossed the shock. Furthermore, how plasma is sampled by PSP throughout the observation introduces a geometrical constraint on the observations, where fluctuations are sampled differently from upstream to downstream. This latter point is further expanded in Figure~\ref{fig:ion_cyclotron}(b) and related discussion. Another crucial caveat stems from the downstream flow speed in the spacecraft frame being much larger than that upstream, allowing more plasma to flow over the spacecraft per unit time, causing the 
downstream SBs to appear shorter than their upstream counterparts, an effect that appears in addition to the compression induced by the shock. Furthermore, the field increase due to the presence of the shock influences the  discussion of the downstream SBs deflection amplitude (Figure~\ref{fig:shockback1}(i)).

For this reason, we decided to ``unshock'' the magnetic field and plasma time 
series across the PSP shock crossing. This analysis is carried out in two steps.
First, using the mean upstream and downstream flow speeds, we plot the PSP
measurements in units of distance from shock using the Taylor 
hypothesis~\citep{Taylor1938}. Then, we used the Rankine-Hugoniot jump conditions to
remove the shock compression. Here, knowing the downstream quantities and the shock 
parameters as estimated from the spacecraft crossing (see 
Table~\ref{tab:tab_event}), we derived the upstream quantities, i.e., we 
estimated the magnetic field and plasma conditions prior to the shock passage. Further details about this method of decompression can be  found in the Appendix~\ref{sec:appendix}.
The resulting, decompressed timeseries is shown in Figure~\ref{fig:decompression}.
As can be seen in panels (a)-(d), the jump in the magnetic field and speed is 
greatly reduced with this technique. It is possible to see that, some compression in magnetic field persists even after the plasma has been ``unshocked''. This is due to the decompression method having several limitations. First, the shock parameters are assumed to be fixed throughout the decompression, and are estimated using a single spacecraft technique, inherently affected by uncertainties~\citep{Trotta2022b}. The shock variability, particularly important 
for supercritical shocks, on a variety of 
scales~\citep[e.g.][]{Marghitu2017,Kajdic2019,Kajdic2021} may therefore introduce fluctuations in the decompression. Finally, as discussed in the appendix, the decompression method makes use of the Rankine-Hugoniot jump conditions without including waves and/or turbulence~\citep[e.g.,][]{Zank2002,Gedalin2023}. This technique can be further improved, in 
future studies, to mitigate the above limitations.

The preservation of the SBs Alfv\'enicity is evident from
Figure~\ref{fig:decompression} (a)-(d),
where the downstream behaviour of the 
magnetic field and velocity fluctuations is remarkably similar to the upstream one. 
The analysis of magnetic field excursions further strengthens this point, where 
compatible behaviour between downstream and upstream fluctuations is found.
The magnetic field excursion space covered by the downstream fluctuations is still 
larger, due to some compressive features downstream of the shock not fully removed 
by the Rankine-Hugoniot decompression. However, we observe that downstream of the 
shock we find larger deflections of the magnetic field. It is also worth noting that,
after decompression,  the difference in the $\mathrm{B_R}$-$\mathrm{B_N}$ plane 
becomes less evident.


A totally different behaviour from that found for SBs is found for the upstream ion cyclotron waves 
discussed in Section~\ref{subsec:event_overview}. The signature of these waves, 
a clearly visible feature in the reduced magnetic helicity spectrogram 
(red patches of $\sigma_m = 1$, near ion plasma frequency, 
Figure~\ref{fig:wav_turb}), is completely lost 
downstream of the shock. We address this interesting feature 
in Figure~\ref{fig:ion_cyclotron}, where we study how PSP
crosses the wave environment surrounding the shock.
To this end, we compute the angle between the 
local magnetic field and the radial direction, $\theta_{\mathrm{BR}}$, and the 
angle between the local magnetic field
and the local velocity $\theta_{\mathrm{BV}}$,
where $\mathbf{V} = \mathbf{V} - \mathbf{V_{PSP}}$,
where $\mathbf{V}_{\mathbf{PSP}}$ is the velocity of PSP in the inertial RTN frame. In Figure~\ref{fig:ion_cyclotron}(b) ion cyclotron waves are clearly observed when \tbv is close to 180 degrees and lost when \tbv is of the order of 90 degrees. This behaviour could be simply due to an observational bias. As shown by ~\cite{BowenICWTurb2020ApJS..246...66B}, this phenomenon is the result of measuring a quasi-parallel wave vector at oblique angles combined with the higher amplitude in the perpendicular direction of the anisotropic turbulent fluctuations.
Furthermore, the conditions downstream of the shock inhibit the possibility of 
resonant beam-field interaction, due to the strongly turbulent environment diffusing
the beam in velocity space~\citep[e.g.,][]{Valentini2010}. We analysed the ion
velocity distribution functions as 
measured by SPANi, albeit their being complicated by the limited field of view of the 
instrument, we did not find any beam signature downstream of the shock, suggesting 
that the production mechanisms of ion cyclotron waves are indeed suppressed. A 
detailed investigation of such features is out of scope for this work and 
will be the object of further studies.

\subsection{The shock at Solar Orbiter} \label{subsec:solo}

In this Section, we discuss the event as observed by SolO. The shock 
crossed the spacecraft at 10:00:51 on Sep 6, 2022. The shock parameters, reported 
in Table~\ref{tab:tab_event}, are compatible with the PSP observations. However, 
as discussed in Section~\ref{subsec:event_overview}, shock variability
plays a major role in the SolO observations. Here, the shock transition appears far more structured, with an overall behaviour compatible with strong shocks.

Such interesting aspects of the SolO shock observations are investigated in 
Figure~\ref{fig:solo_tbn}. The structuring observed in the shock surroundings 
is particularly evident in the 20 minutes downstream of the shock, populated with many sharp changes of magnetic field (Figure~\ref{fig:solo_tbn}). 
First of all, using the EPD-EPT instrument, we 
address how the production of energetic particles is affected by such disturbed conditions. As can be seen in 
Figure~\ref{fig:solo_tbn}(a), energetic ions of up to $\sim$ 2 MeV are 
accelerated by the shock, with fluxes rising as the shock crossing is approached, a behaviour typical of ion acceleration at IP shocks~\citep{Giacalone2012,Lario2022}. 
Interestingly, some of the fluxes (e.g. the ones for $\sim$0.33 and $\sim$ 
0.7 MeV, green and yellow lines, respectively) have their peak at a time 
which is not coincident with the shock crossing time at SolO, but instead  
is downstream of the shock. This behaviour, consistent with the statistical study presented in \citet{Lario2003}, has implications on the production 
and propagation of shock accelerated particles~\citep{Perri2022}, and may well be due to additional acceleration mechanisms downstream of the shock~\citep{Zank2015, Zhao2018, Kilpua2023}.

A crucial feature found for the energetic particle population is the 
fluctuations found in their fluxes immediately downstream of the shock with 
typical timescales of 1 to 2 minutes, 
particularly evident for the lower energy channels (0.05 to 0.35 MeV, dark 
blue to green). Such
fluctuations correlate well with 
the magnetic field structuring (Figure~\ref{fig:solo_tbn}(a)-(c)), thus 
suggesting that particle acceleration is indeed happening in an irregular fashion 
for this IP shock, where additional acceleration may be provided by the magnetic structures~\citep{Zhao2018,Trotta2022a,Nakanotani2021}. To quantify the variability associated with the shock 
crossing, the \tbn and local \tbr angles have been estimated using the local 
magnetic field, the radial direction and the average shock normal as 
reported in Table~\ref{tab:tab_event}. It is evident from this 
analysis, shown in in Figure~\ref{fig:solo_tbn}(b), that the shock geometry 
changes significantly on the timescales examined here, moving from quasi-perpendicular
to quasi-parallel geometries and thus supporting the picture 
of irregular particle acceleration. 
Complementing the theme of shock variability, it is worth discussing the 
role of plasma structures in particle acceleration. The typical gyroradii of 
the particles showing such irregular behaviour have been estimated 
using the local magnetic field, $\mathrm r_L \sim 10^4$ km. Using the Taylor's 
hypothesis, such lengths compare to 1/10th of the typical size of the 
downstream structures, thus suggesting that trapping of accelerated 
particles may play a role into the variability observed in the particle 
fluxes, an important source of extra particle acceleration beyond the 
shock~\citep[see][]{Zank2015, Zhao2018,Trotta2020b, LeRoux2019, Pezzi2022}. 

\begin{figure}
	\includegraphics[width=.5\textwidth]{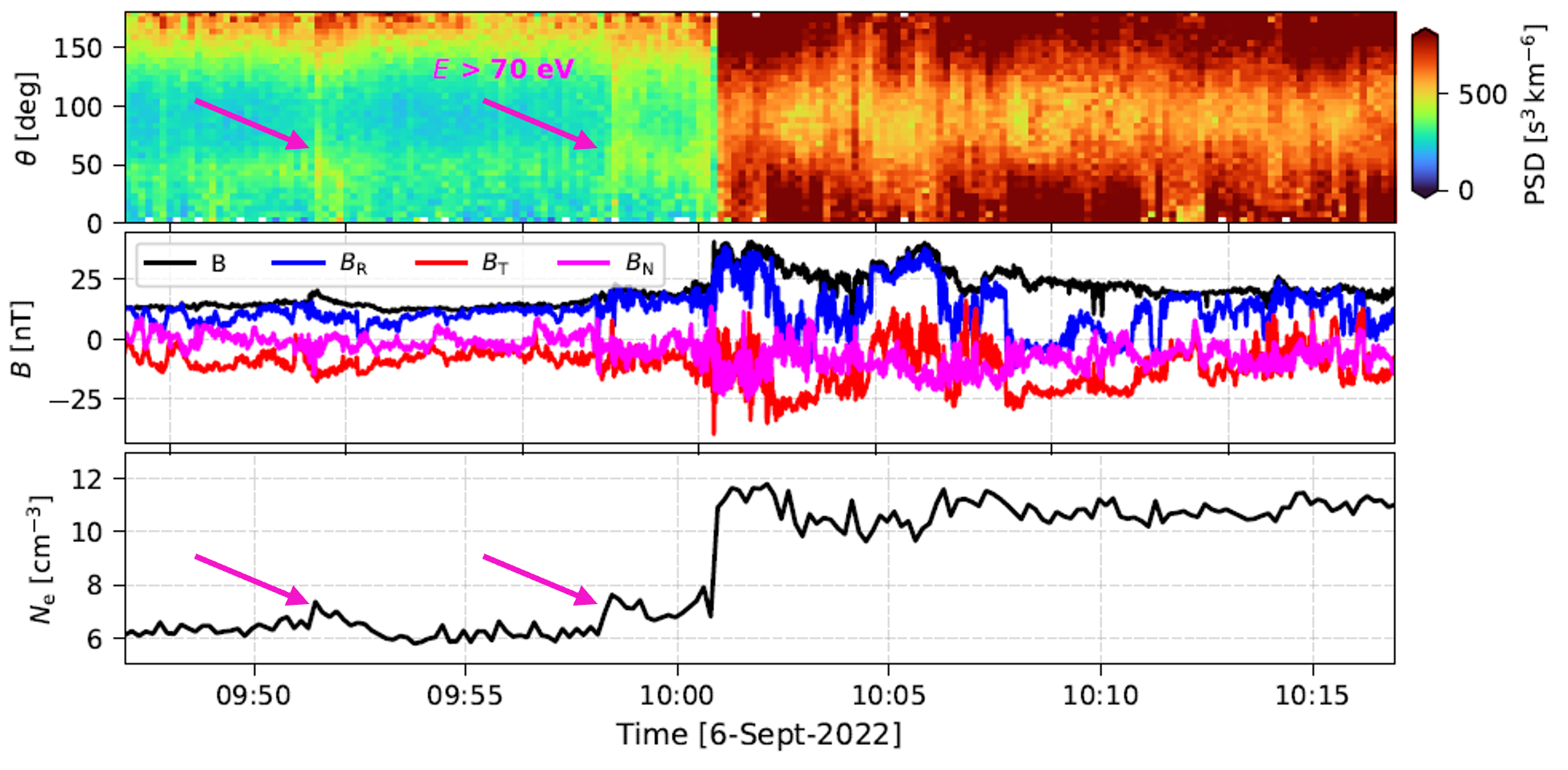}
    \caption{Overview of electron behaviour at SolO. Top: EAS electron pitch angle distributions for particles with energy larger than 70 eV. Middle: Magnetic field measurements in RTN.Bottom: electron density computed from EAS measurements.}
    \label{fig:eas_shocklets}
\end{figure}

Another extremely interesting feature of the SolO observations is the steep 
magnetic field enhancements observed in the 10 minutes upstream of the 
shock (Figure~\ref{fig:eas_shocklets}). 
Such steep enhancements are compatible with upstream shocklets, 
typically arising from the interplay between shock-reflected, energetic ions 
and upstream waves which, in the case of Earth's bow shock, are the Ultra Low 
Frequency (ULF) waves routinely found in the foreshock~\citep{Stasiewicz2003}.
Shocklets are important as 
they influence the plasma environment upstream of the shock transition, thus 
pre-conditioning the shock inflow~\citep{Lucek2008}. Most of our 
knowledge of shocklets is built on observations upstream of Earth's bow 
shock~\citep{Plaschke2018}, being rarely observed at 
IP shocks with only three cases previously 
reported~\citep{Lucek1997,Wilson2009,Trotta2023a}, making this observation particularly interesting.

Two shocklets were observed upstream of the shock at SolO, compatible with the 
fact that, when estimated at $\sim$ 20 minute timescales, the shock Mach number is 
high and the geometry quasi-parallel (see discussion in 
Section~\ref{subsec:event_overview}). In Figure~\ref{fig:eas_shocklets}, we show 
how the shocklets influence the low-energy electron population using SolO-EAS 
measurements. First of all, shocklets are associated with rises in the electron 
density, identified by the magenta arrows in Figure~\ref{fig:eas_shocklets}(c), a behaviour recently found also at Venus~\citep{Collinson2023}.
Furthermore, by analysing the pitch-angle distributions of electrons with energy 
greater than 70 eV, we found that electrons are more isotropic in pitch angle 
space in the vicinity of the shocklets, probably due to pitch-angle scattering 
induced by the compressive structures, as observed in the Earth's bow shock 
environment~\citep{Wilson2013}. This behaviour, namely the pre-conditioning of 
the upstream particle population, may have 
important consequences for electron injection and acceleration to higher 
energies~\citep{Katou2019, Dresing2022, Wijsen2023}.

\section{Conclusions}\label{sec:conclusions}

The September 05, 2022 CME event provided an opportunity to study a CME-driven IP shock  at unprecedentedly low heliocentric 
distance, with PSP crossing the event at 0.07 au. Furthermore, well-radially 
aligned SolO at 0.7 au provided important insights about the 
evolution of the event. 

We focused on the small-scale behaviour of CME-driven shocks associated
with this event. The PSP spacecraft observed the shock very early in its 
evolution. The shock had moderate Mach numbers
($\mathrm{M_A} \sim \mathrm{M_{fms}} \sim 4)$), as inferred  
using local shock parameter estimation. The average shock normal was 
found to be $\mathrm{\hat{n}} = (0.5, -0.8, 0.2)$, significantly departing 
from the radial direction, indicating that the crossing of the shock likely
happened on one flank of the CME event. This has also important implications when addressing the joint PSP-SolO observations, as due the shock will be crossed in different loactions, with features depending on the CME width and spacecraft longitudinal separation. The shock at PSP has a notably
short sheath region due to its early evolutionary stage, an interesting property 
that may have fundamental implications for 
the possibility of accelerating particles to high energies, as preliminary shown in~\citet{Cohen2023}. We however emphasize that as this event was crossed at the flanks of the sheath - CME flux rope boundary cannot be determined accurately. 

We studied how switchbacks, fundamental constituents of the solar wind 
as elucidated by several, previous PSP observations~\citep[e.g.][]{Bale2019, kasper19, DudokdeWitSBs2020},
are processed in the shock crossing. We found that the SBs are compressed 
along the shock normal direction and stretched along the other two directions. 
Interestingly, many SBs properties are preserved in the shock downstream. 
Furthermore, SBs with larger magnetic field deflections were found in the 
shock  downstream, an important ingredient to consider when addressing 
their statistical properties. Such statistical properties of how SBs are processed by IP shocks will be object of future investigation, now increasingly possible thanks to the novel PSP ShOck Detection Algorithm (SODA) IP shock list \url{https://parker.gsfc.nasa.gov/shocks.html}. This behaviour has been investigated in 
detail by looking at the decompressed magnetic field and plasma quantities,
obtained using the Rankine-Hugoniot shock jump conditions. 
This technique of decompression is useful to estimate how quantities 
were before their interaction with the shock. It can also be used to 
yield an estimate of how the ambient plasma conditions
could change due to the passage of a shock with a given set of parameters,
as will be shown in future work.

How ion cyclotron waves are transmitted across the shock was also addressed.
This study is particularly interesting due 
to the role that ion cyclotron waves play for energy 
dissipation~\citep[e.g.,][]{Bowen2022}. 
We observe that the signature of pre-existing 
ion cyclotron waves, identified in the shock upstream at PSP, disappears downstream
of the shock (Figure~\ref{fig:ion_cyclotron}). This may be due to 
the sudden change in plasma conditions at the shock, injecting strong 
fluctuations in the downstream, making the conditions for ion cyclotron waves 
propagation less favourable. Through the analysis of the \tbv and \tbr angles,
we also found unfavourable conditions to detecting ion cyclotron activity 
downstream~\citep{BowenICWTurb2020ApJS..246...66B}, due to the change of the mean magnetic field 
direction upon the shock passage. 

SolO observations of the same event are extremely interesting 
to address the role of evolution for the CME shock region in its propagation
to larger heliocentric distances. On the large scales,
we note that the large transverse flow deflections are still present,
with a $\mathrm{V_N}$ increase comparable to the one observed in 
$\mathrm{V_R}$.

The shock environment at SolO is much more disturbed than the one observed 
at PSP. A shock parameter estimation using very short averaging windows ($\sim$ 1
minute, thus addressing the very local shock properties) yields similar 
values with respect to the PSP observations (see Table~\ref{tab:tab_event}). 
However, analysing the 30 minutes across the shock transition, we find an
environment compatible with a quasi-parallel shock and relatively high
Mach number, propagating in a very structured portion of the solar wind.
Two shocklets, structures that grow favourably upstream of high Mach number,
quasi-parallel shocks, were found upstream of the shock, a rare observation for
the interplanetary case. Signature of $\sim 100$ eV electrons scattering was found 
corresponding to the shocklets, an important ingredient to be considered for
pre-conditioning of the upstream particle population at the shock. It is worth 
noting that these shocklets, with duration much larger than those observed at 
Earth's bow shock, probably arise from pre-existing upstream waves and not from 
shock-generated upstream waves, a behaviour observed also for the case in 
~\citet{Trotta2023a}. Therefore, this observation points to the idea that the 
origin of the IP shocklets is different than those observed in the Earth's 
foreshock.

Energetic ions up to 2 MeV were found in association with the shock at SolO. 
A detailed analysis of the high time resolution EPD-EPT energetic 
particle fluxes reveal structuring corresponding to the magnetic field structures 
processed by the shock, indicating a potential role of trapping as 
an extra source of energy for particle acceleration. 

Finally, we underline that this event is a very good example 
of the novel observational window provided by missions exploring 
the inner heliosphere such as PSP and SolO, exploited in this study  to highlight the fact that the features of IP shock 
environments can be very different as they propagate through the heliosphere, 
with important consequences on the modelling effort accuracy and possibility of
prediction associated with such energetic events.


\section{acknowledgments}
This work has received funding from the European Unions Horizon 2020 research 
and innovation programme under grant agreement No. 101004159 (SERPENTINE, 
www.serpentine-h2020.eu). AL is supported by STFC Consolidated Grant 
ST/T00018X/1. CHKC is supported by UKRI Future Leaders Fellowship MR/W007657/1 
and STFC Consolidated Grants ST/T00018X/1 and ST/X000974/1.
H.H. is supported by the Royal Society University Research Fellowship URF/R1/180671. N.D. is grateful for support by the Academy of Finland (SHOCKSEE, grant No.\ 346902). The PSP/FIELDS experiment was developed and is operated under NASA contract NNN06AA01C. Work at IRAP is supported by CNRS, UPS, and CNES.E.E.D acknowledges funding by the European Union (ERC, HELIO4CAST, 101042188). Views and opinions expressed are however those of the author(s) only and do not necessarily reflect those of the European Union or the European Research Council Executive Agency. Neither the European Union nor the granting authority can be held responsible for them. XBC is grateful to PAPIIT DGAPA grant IN110921. This work was supported by the UK Science and Technology Facilities Council (STFC) grant ST/W001071/1.L.F. .is supported by the Royal Society University Research Fellowship No. URF\textbackslash R1\textbackslash231710. RFWS thanks the
German Space Agency (DLR) for its support of SOLO/EPD under grant 50OT2002.

%






\appendix
\section{Rankine-Hugoniot decompression method}
\label{sec:appendix}

Heliospheric shocks crossing spacecraft can be directly measured. An hypothesis that is often made when interpreting spacecraft measurements is the Taylor hypothesis~\citep{Taylor1938}, linking time-variations in the spacecraft measured quantities to spatial variations. Thus, upon shock crossing, it is possible to address the properties of the upstream/downstream shock environments (see Figure~\ref{fig:overview}). However, it is often the case that it would be interesting to address the shock downstream plasma \textit{before it was shocked}, as discussed in Section~\ref{subsec:shockback}. Here we present a technique providing a proxy to address the plasma condition before the shock propagation.

\begin{figure*}
	\includegraphics[width=\textwidth]{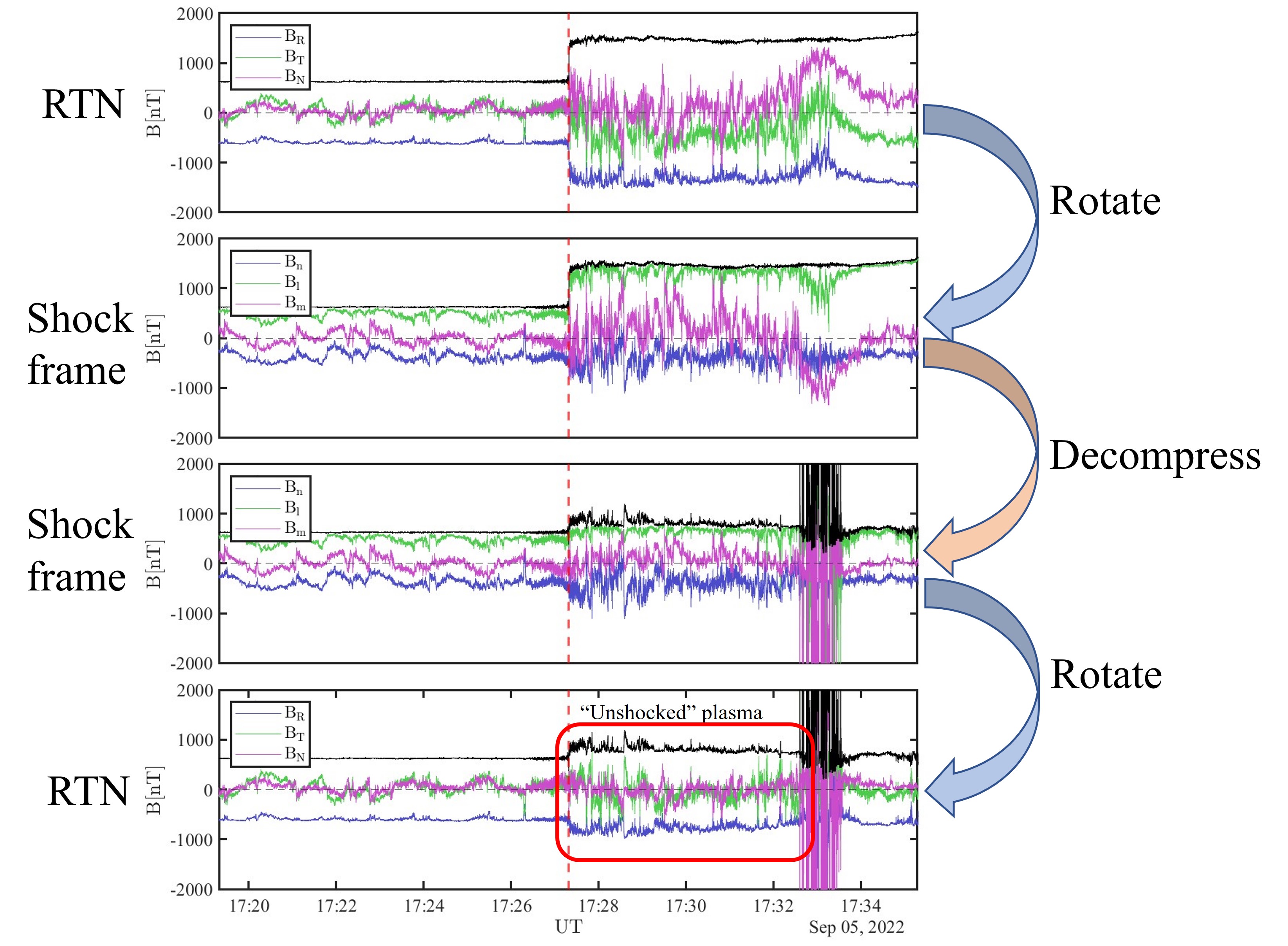}
	\caption{Overview of the procedure of magnetic field decompression.}
	\label{fig:appendix_derankine}
\end{figure*}

The Rankine-Hugoniot jump conditions \citep[e.g.,][]{Burgess2015} have been widely used to model the plasma properties across a shock. Here, the shock is treated as a planar, one-dimensional, time stationary structure. Following an integration of the MHD equations for mass, momentum and energy conservation, and the divergence-free condition for the magnetic field, the Rankine-Hugoniot relations linking the properties of upstream and downstream plasmas can be written as follows:

\begin{align}
	\frac{\rho_{m2}}{\rho_{m1}} &= \mathrm{r_{gas}} \label{eq:o1} \\
	\frac{V_{n2}}{V_{n1}} &= \frac{1}{\mathrm{r_{gas}}} \\
	\frac{V_{t2}}{V_{t1}} &= \frac{M^2_{A1} - 1}{M^2_{A1} - \mathrm{r_{gas}}} \\
	\frac{B_{n2}}{B_{n1}} &= 1 \label{eq:o4} \\
	\frac{B_{t2}}{B_{t1}} &= \mathrm{r_{gas}}\frac{M^2_{A1} - 1}{M^2_{A1} - \mathrm{r_{gas}}} \label{eq:o5} \\
	\frac{P_2}{P_1} &= \mathrm{r_{gas}} + \frac{(\gamma -1) \mathrm{r_{gas}} V_1^2}{2V_S1}\left( 1 - \frac{V_2^2}{V_1^2}\right) \label{eq:o6}.
\end{align}

The above equations for density, bulk flow velocity, magnetic field and pressure ($\rho_m, V,B$ and $P$, respectively) are expressed in the deHoffmann-Teller frame, i.e., a frame aligned to the shock normal that moves at a speed such that the upstream convective electric field ($\mathbf{E}_1 = - \mathbf{V}_1 \times \mathbf{B}_1$) vanishes \citep{deHoffmann1950}. Here, the subscripts 1 and 2 are referred to the upstream and downstream states, respectively. The subscripts $n$ and $t$, instead, indicate the shock normal and tangential directions. $\mathrm{r_{gas}}$ is the shock gas compression ratio, and $M_{A1} \equiv V_{n1}\sqrt{\mu_0 \rho_{m1}}/B_{n1}$ is the Alfv\'enic Mach number. $V_{S1}$ is the upstream sound speed.

Often, the Rankine-Hugoniot equations are used to address shock parameters. What we do in our decompression technique is instead to use the downstream measurements and the shock parameters to compute the upstream conditions, thus ``unshocking'' the plasma according to Equations~\ref{eq:o1}-\ref{eq:o6}. The procedure, given a time-series of spacecraft measurements, is performed as follows, and displayed in Figure~\ref{fig:appendix_derankine}.
First, the data is rotated in a shock normal frame. Here, we choose the $nlm$ frame, where the $n$ direction is aligned with the shock normal (computed using the MX3 method~\citep{Paschmann2000}), $m$ is perpendicular both to the shock normal and to the upstream magnetic field, and $l$ completes the triad. A boost is performed to then move to the deHoffman-Teller frame. Then, Equations~\ref{eq:o1}-\ref{eq:o6} are used to derive the upstream quantities given the downstream measurements and the shock parameters, i.e., the decompression is performed. The data is then returned in the spacecraft frame. Finally, using the mean upstream and downstream flow speeds, it is possible to shock the measurements in units of distance from shock (see Figure~\ref{fig:decompression}).

This technique, while giving a proxy for the plasma conditions before shock processing, has several limtations. The decompression is performed assuming that the shock parameters do not change throughout the event, thus neglecting the inherent variability of shock system. Shock parameters are estimated from single spacecraft measurements, and therefore are associated to uncertainties. The Rankine-Hugoniot relations used to decompress the plasma assume a one-dimensional, planar, time-stationary MHD shock with laminar upstream/downstream regions, which is notoriously a stringent assumption for heliosphere shocks, that are characterised by several space/time variabilities and propagate in the turbulent solar wind.

For the above reasons, it is readily understood that the decompression is more reliable for closer measurements with respect to the shock crossing. In Figure~\ref{fig:appendix_derankine}, it is possible to see how unphysical results are introduced by the large structure embedded in the plasma happening around 17:33 UT. Future improvements of the diagnostic for example including wave transmission will be object of future work.





\end{document}